\documentclass[unnumsec,webpdf,modern,large,namedate]{oup-authoring-template}% uncomment this line for author year citations and comment the above

\usepackage[dvipsnames]{xcolor}

\graphicspath{{Fig/}}

\begin{document}

\newcommand{\revision}[1]{\textcolor{Red}{#1}}

\journaltitle{arXiv}
\DOI{}
\copyrightyear{2025}
\pubyear{2025}
\appnotes{\textbf{Application Note}}

\firstpage{1}

\subtitle{Gene Expression}

\title[DeeDeeExperiment]{DeeDeeExperiment: Building an infrastructure for integrating and managing omics data analysis results in R/Bioconductor}

\author[1$\ast$]{Najla Abassi\ORCID{0000-0001-8357-0938}}
\author[1]{Lea Schwarz \ORCID{0000-0003-2554-0583}}
\author[1,3]{Edoardo Filippi \ORCID{0009-0003-9858-0137}}
\author[1,2$\ast$]{Federico Marini\ORCID{0000-0003-3252-7758}}

\authormark{Abassi et al.}

\address[1]{\orgdiv{Institute of Medical Biostatistics, Epidemiology and Informatics (IMBEI)}, \orgname{University Medical Center Mainz}, \state{Mainz}, \country{Germany}}
\address[2]{\orgdiv{Research Center for Immunotherapy (FZI) Mainz},\state{Mainz}, \country{Germany}}
\address[3]{\orgdiv{Department of Nephrology, Rheumatology and Kidney Transplantation}, \orgname{University Medical Center Mainz},\state{Mainz}, \country{Germany}}

\corresp[$\ast$]{Corresponding authors. \href{mailto:najla.abassi@uni-mainz.de}{najla.abassi@uni-mainz.de} and \href{mailto:marinif@uni-mainz.de}{marinif@uni-mainz.de}}

\received{Date}{0}{Year}
\revised{Date}{0}{Year}
\accepted{Date}{0}{Year}

%\editor{Associate Editor: Name}

%\abstract{
%\textbf{Motivation:} .\\
%\textbf{Results:} .\\
%\textbf{Availability:} .\\
%\textbf{Contact:} \href{name@email.com}{name@email.com}\\
%\textbf{Supplementary information:} Supplementary data are available at \textit{Journal Name}
%online.}

\abstract{
\textbf{Summary:}
Modern omics experiments now involve multiple conditions and complex designs, producing an increasingly large set of differential expression and functional enrichment analysis results. However, no standardized data structure exists to store and contextualize these results together with their metadata, leaving researchers with an unmanageable and potentially non-reproducible collection of results that are difficult to navigate and/or share. Here we introduce DeeDeeExperiment, a new S4 class for managing and storing omics data analysis results, implemented within the Bioconductor ecosystem, which promotes interoperability, reproducibility and good documentation. This class extends the widely used SingleCellExperiment object by introducing dedicated slots for Differential Expression (DEA) and Functional Enrichment Analysis (FEA) results, allowing users to organize, store, and retrieve information on multiple contrasts and associated metadata within a single data object, ultimately streamlining the management and interpretation of many omics datasets. \\
\textbf{Availability and implementation:}
DeeDeeExperiment is available on Bioconductor under the MIT license (\url{https://bioconductor.org/packages/DeeDeeExperiment}), with its development version also available on Github (\url{https://github.com/imbeimainz/DeeDeeExperiment}).
}

% \boxedtext{
% \begin{itemize}
% \item Key boxed text here.
% \item Key boxed text here.
% \item Key boxed text here.
% \end{itemize}}

\maketitle

\section{Introduction}
Over the past decade, omics technologies have advanced the study of human health and diseases, as well as a wide range of model organisms generating data at unprecedented scale \citep{Smirnov2023, Chen2023, VandenBerge2019}. As the volume of these data grows, so does the challenge of analyzing and managing the increasingly heterogeneous results they produce. In practice, many of these results take the form of tables generated from differential expression analysis (DEA) and functional enrichment analysis (FEA), such as Over-Representation Analysis and Gene Set Enrichment Analysis \citep{Ritchie2015, Love2014, Robinson2009, Chen2013, Yu2012, Alexa2006, Korotkevich2021}. These steps are central in most omics workflows, enabling researchers to detect and interpret biological differences between conditions. However, as the number of conditions and analysis tools increases, these outputs can quickly accumulate in large numbers, making their management overwhelming even for expert practitioners. This challenge is evident in bulk RNA-seq and other omics modalities (e.g., proteomics or metabolomics), where complex experimental designs can generate extensive results \citep{Peng2024, Hajjar2023}. It becomes even amplified in single-cell RNA-seq studies, where the pseudobulk analysis (i.e., aggregated at the cluster-sample level) yields numerous results tables across multiple conditions/cell types \citep{Crowell2020}.

Despite the availability of numerous DEA and FEA frameworks implemented within the Bioconductor project, which relies on interoperable S4 classes for omics analyses,there is currently no standardized data structure for organizing, linking, and contextualizing multiple DEA and FEA results together with their associated metadata. Without such infrastructure, it becomes difficult to keep track of the analytical context, recall parameters, or share results in a reproducible manner, particularly in collaborative settings.

Here we describe DeeDeeExperiment, a new S4 class that extends the widely adopted SingleCellExperiment \citep{Amezquita2019} Bioconductor object. The class provides a standardized data structure for organizing DEA and FEA results alongside expression data and metadata. Our framework enables users to efficiently retrieve and explore outputs across multiple contrasts in a structured and reproducible manner. By building it within the Bioconductor ecosystem, DeeDeeExperiment leverages its long-standing principles of interoperability, continuous testing, optimization, and comprehensive documentation \citep{Huber2015}, ensuring compatibility with existing downstream tools and workflows, including those for interpretation and interactive exploration \citep{Marini2021, RueAlbrecht2018, Ludt2022}.

\section{Implementation}\label{sec2}
\begin{figure*}[!h]%
\centering
\includegraphics[width=0.95\linewidth]
{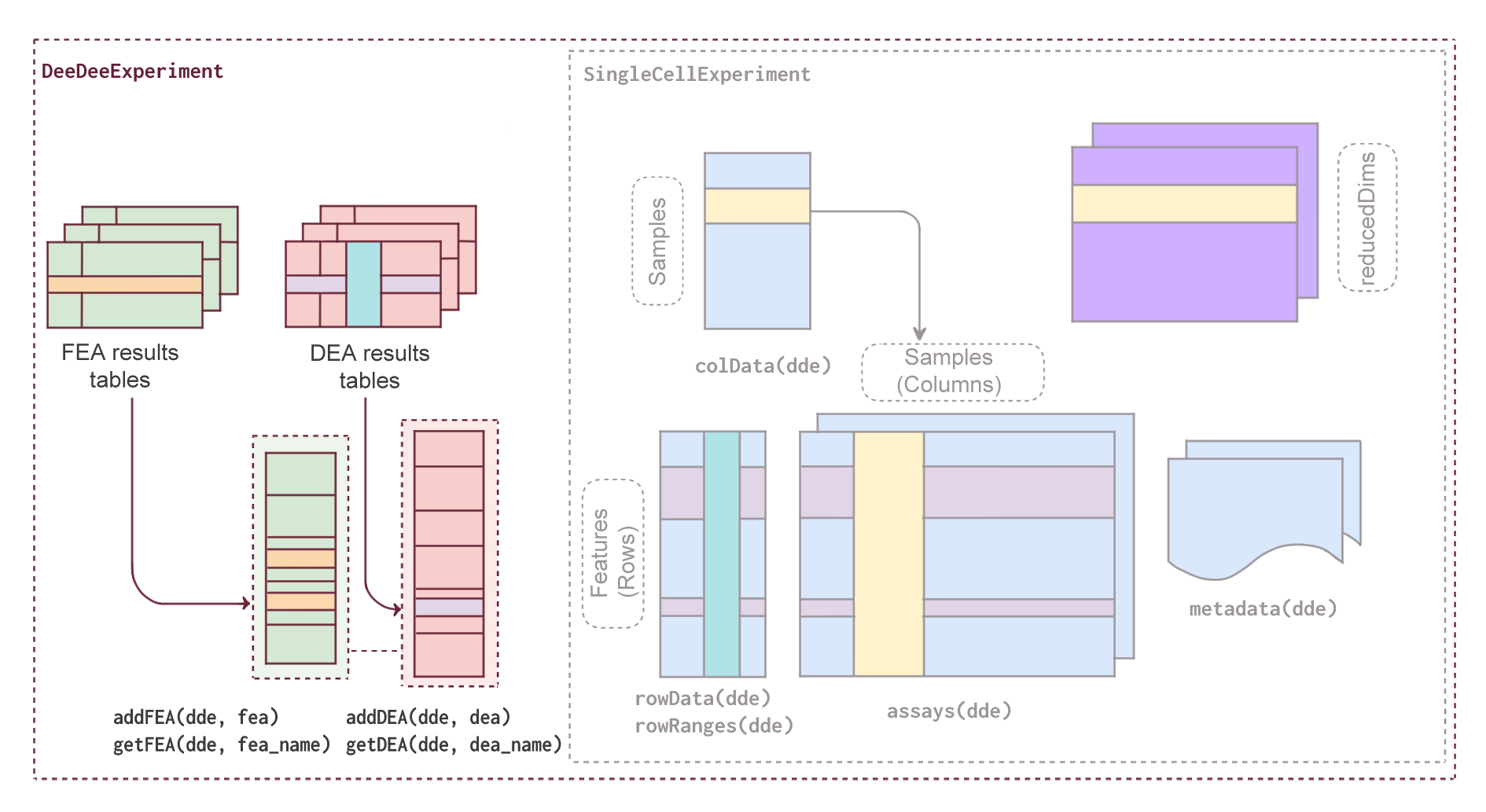}
\caption{Anatomy of DeeDeeExperiment class. It extends the SingleCellExperiment class (right panel), inheriting its core slots (\texttt{assays}, \texttt{rowData}, \texttt{colData}, \texttt{reducedDims}, \texttt{metadata}) and adds two new slots: \texttt{dea}, for storing Differential Expression Analysis (DEA) results, and \texttt{fea}, for storing Functional Enrichment Analysis (FEA) results. Both slots store results as a list of named contrasts/comparisons (hereby used interchangeably). Their names are provided by the user (defaulting to the object name if not specified), and are enforced to be unique. For each contrast, the \texttt{dea} slot stores (when available) metadata describing the DEA, such as the alpha level, LFC threshold, package used to generate the results and its version. The \texttt{fea} slot follows the same design. For each contrast, it stores metadata such as the associated DEA contrast from which the FEA was derived, the FEA name, the "shaken" GeneTonic output, the tool used to perform the enrichment analysis and its version. Original result objects and feature-level summaries are linked through the metadata and rowData.}
\label{fig1}
\end{figure*}

The DeeDeeExperiment package includes the class definition, along with methods to create, access, and manipulate DeeDeeExperiment objects. Objects can be created with the constructor function \texttt{DeeDeeExperiment()}.
Since the class inherits from SingleCellExperiment, it retains all its core slots, including: (i) \texttt{assays}, a matrix-like slot (e.g. a matrix or DelayedArray) that holds experimental data such as raw counts and/or normalized values across a common set of samples or cells, (ii) \texttt{rowData}, a DataFrame object describing the features under inspection (e.g. gene identifiers and all the associated information), (iii) \texttt{colData}, a DataFrame describing the samples or cells (such as experimental condition, batch, donor, cell type annotation, or quality control metrics), (iv) \texttt{metadata}, for additional information,  and (v) the \texttt{reducedDims}, which stores reduced-dimensional representations of the data, such as PCA, t-SNE or UMAP \citep{Jolliffe2016, maaten2008visualizing, https://doi.org/10.48550/arxiv.1802.03426}.
In addition, DeeDeeExperiment introduces two new slots: (vi) \texttt{dea}, which stores Differential Expression Analysis (DEA) results as standard objects from DESeq2 \citep{Love2014}, edgeR \citep{Robinson2009} and limma \citep{Ritchie2015} or a data.frame provided by the user that contains at least three required columns: log2FoldChange, pvalue, and padj. Additionally, the (vii) \texttt{fea} slot stores Functional Enrichment Analysis (FEA) results from commonly used packages such as topGO \citep{Alexa2006}, clusterProfiler \citep{Yu2012}, or GeneTonic \citep{Marini2021} among others (Fig. \ref{fig1}). The \texttt{dea} and \texttt{fea} slots are implemented as a set of named contrasts, each carrying basic metadata such as the package and version used to generate the results. To ensure efficient storage, the full original DEA results objects for each comparison are stored in the metadata slot of DeeDeeExperiment. Each contrast entry retains an internal pointer referencing its associated original object, allowing users to retrieve it seamlessly when needed. Feature-level DEA statistics (logFC, p-value, and adjusted p-value) for each contrast are embedded into the rowData slot of DeeDeeExperiment.

Our implementation provides methods to add, remove, and retrieve stored analysis results for different contrasts, as well as a \texttt{summary()} method for inspecting them. The summary method provides a quick contrast-level overview, such as the number of up/down-regulated genes or enriched terms.
To support multi-condition workflows, DeeDeeExperiment provides helper functions for integrating results from limma objects with multiple contrasts \citep{Ritchie2015} and pseudobulk analyses generated with muscat \citep{Crowell2020}. As described above, the original DEA results objects are stored once within the metadata and referenced across contrasts. For limma, this means the underlying MArrayLM object is linked rather than duplicated. In contrast, muscat pseudobulk workflows produce separate results tables per contrast, which are stored individually but still organized coherently via the same metadata infrastructure.

Existing common practices based on the original Summarized\-Experiment or SingleCellExperiment objects \citep{https://doi.org/10.18129/b9.bioc.summarizedexperiment, Amezquita2019} combined with ad hoc result lists do not yet provide a unified, contrast-centric container that co-stores DEA and FEA results and integrates feature-level statistics into the rowData across multiple contrasts. Other packages also introduce domain-specific data structures, for example, MultiAssayExperiment for multi-omics integration \citep{Ramos2017}, or the DatasetExperiment class in structToolbox \citep{Lloyd2020} for metabolomics. However, these solutions are designed to organize assay-level data rather than providing a general, contrast-centric S4 framework for storing and managing multiple DEA and FEA analysis results. Our proposed class, DeeDeeExperiment, fills this niche while preserving seamless compatibility with most other Bioconductor tools.

To further enhance reproducibility, contextual information related to specific DEA results can be stored in a DeeDeeExperiment object using the \texttt{addScenarioInfo()} method. This functionality helps document the experimental setup, clarify comparisons, and facilitate downstream interpretation. Looking ahead, such contextual metadata could also be leveraged by large language models (LLMs) to enhance machine-assisted interpretation \citep{Hao2024}, opening opportunities for integrating DeeDeeExperiment objects with emerging AI-driven tools.
Since DeeDeeExperiment extends the SingleCellExperiment class, it maintains compatibility with the broad ecosystem of Bioconductor tools. We envision extending compatibility to other classes derived from SingleCellExperiment, such as SpatialExperiment \citep{Righelli2022} and TreeSummarizedExperiment \citep{Huang2021} among others, which do not directly inherit from DeeDeeExperiment. To support this, coercion methods could be implemented to allow users to leverage DeeDeeExperiment's full functionality while preserving class-specific slots. In practical settings, this means that once an analysis is completed and a DeeDeeExperiment object is shared, collaborators can readily visualize the data with scater \citep{McCarthy2017} (e.g., tSNE or UMAP) or explore it interactively with iSEE \citep{RueAlbrecht2018} without performing any extra data handling. A complete documentation and description of use cases for the DeeDeeExperiment package is provided in its vignette (rendered at https://imbeimainz.github.io/DeeDeeExperiment/).

\section{Conclusion}\label{sec3}
DeeDeeExperiment provides a robust, standardized and extensible class for managing, and integrating DEA and FEA results. By storing these results in a conceptually clear manner and capturing the essential details of each analysis within a single container, DeeDeeExperiment enables users to organize, retrieve, explore, contextualize, and interpret their analysis results more efficiently across multiple contrasts, or even across multiple experimental scenarios. Consolidating all information in one data object supports more nuanced and quantitative approaches beyond simple overlap strategies, such as Venn diagrams, enabling trustworthy summaries of complex experimental measurements.

While DeeDeeExperiment is designed to store a wide set of DEA and FEA results, future work will be required to capture additional provenance details such as sessions and environments, ensuring an even higher standard of reproducibility.
Being part of the Bioconductor ecosystem, DeeDeeExperiment maintains full compatibility with other existing tools for downstream analyses, promoting reproducibility and elevating individual DE/enrichment results and their metadata as essential component of research output to share not only among collaborators, but also to the broader scientific community.
DeeDeeExperiment is available on Bioconductor (\url{https://bioconductor.org/packages/DeeDeeExperiment}), and the development version is maintained on GitHub (\url{https://github.com/imbeimainz/DeeDeeExperiment}).

\section{Competing interests}
No competing interest is declared.

\section{Author contributions statement}
N.A. Conceptualization, data curation, methodology, software, validation, writing – original draft, writing – review \& editing
L.S. Data curation, methodology, software, writing – review \& editing
E.F. Validation, writing – review \& editing
F.M. Conceptualization, methodology, software, supervision, funding acquisition, project administration, resources, writing – review \& editing.
All authors approved the final version of the manuscript.

\section{Acknowledgments}
This work was supported by the Deutsche Forschungsgemeinschaft (DFG, German Research Foundation) Projektnummer 318346496 - SFB1292/2 TP19N (to NA and FM).
The authors thank the Bioconductor community for valuable feedback and suggestions, and also thank Ahmed Hassan, Annekathrin Nedwed, Alicia Schulze, and Alina Jenn for the insightful discussions and constructive suggestions on features and functionality, as well as for testing early versions of the package.

\bibliographystyle{abbrvnat}
\bibliography{references_DeeDeeExperiment}

\end{document}